\pdfoutput=1

\documentclass[11pt]{article}

\usepackage[final]{naacl2021}

\usepackage{times}
\usepackage{latexsym}

\usepackage[T1]{fontenc}

\usepackage[utf8]{inputenc}

\usepackage{microtype}

\usepackage{graphicx}

\usepackage{tikz}
\usepackage{xcolor}

\title{Submission-Aware Reviewer Profiling for Reviewer Recommender System}

\author{Omer Anjum \\
  University of Texas, Houston\\
  \texttt{anjum.omer@gmail.com} \\\And
  Alok Kamatar \\
  University of Illinois Urbana-Champaign \\\AND
  Toby Liang \\
  University of Illinois Urbana-Champaign \\\And
  Jinjun Xiong \\
  University at Buffalo \\\AND
  Wen-mei Hwu \\
  University of Illinois Urbana-Champaign}

\begin{document}
\maketitle
\begin{abstract}
Assigning qualified, unbiased and interested reviewers to paper submissions is vital for maintaining the integrity and quality of the academic publishing system and providing valuable reviews to authors. However, matching thousands of submissions with thousands of potential reviewers within a limited time is a daunting challenge for a conference program committee. Prior efforts 
based on topic modeling 
have suffered from losing the specific context that help define the topics in a publication or submission abstract. Moreover, in some cases, topics identified are difficult to interpret. We propose an approach that learns from each abstract published by a potential reviewer the topics studied and the explicit 
context in which the reviewer studied the topics. Furthermore, we contribute a new dataset for evaluating reviewer matching systems.  Our experiments show a significant, consistent improvement in precision when compared with the existing methods. We also use examples to demonstrate why our recommendations are more explainable. The new approach has been deployed successfully at top-tier conferences in the last two years.
\end{abstract}

\section{Introduction}\label{sec:intro}
The peer-review process, which subjects scholarly papers to the scrutiny of experts in the area, is essential for controlling the quality of scientific publications. A critical step of the process is to identify qualified reviewers for each new submission. 
However, the traditional manual process of assigning reviewers is under immense strain due to the record-breaking, ever-increasing number of submissions \cite{subtrends, acl_acceptance}.

For instance, the ISCA 2018 Program Committee reflected, \textit{``... there was disparity in how authors chose topic keywords for submissions with many only using a single keyword and others using over half a dozen keywords. As such, relying on the key-words for submissions is difficult. ... So, we chose to hand assign the papers''} \cite{sigarchBlog}. And again in 2019, the ACL Program Committee stated, \textit{``Our plan was to rely on the Toronto Paper Matching System(TPMS) ... this system did not prove as useful as we had hoped for (it requires more extensive reviewer profiles for optimal performance ... the work had to rely largely on the manual effort...''}\cite{aclBlog}.
Among the prior methods for reviewer recommendation proposed by NLP community, bag-of-words based models have been widely used, mainly Latent Semantic Indexing (LSI) \cite{LSI} and Latent Dirichlet Allocation (LDA) \cite{LDA}. To learn meaningful representations of a reviewer, both LSI and LDA require a very large corpus to learn hidden relations between words and documents. Since the relations are captured in a global context, the resulting reviewer representation can be too diluted to be meaningfully compared to the main focus of any particular submission.

Some other approaches \cite{hiddentopic,pare} use vector embeddings \cite{w2v}. To capture underlying topics from text, these approaches calculate a low-rank approximation \cite{svd} of the embedding matrix. It can be helpful to overcome the potential vocabulary mismatch between a submission abstract and abstracts of the papers previously published by a reviewer (referred to as reviewer abstracts for the rest of this paper). However, these approaches still do not capture the specific local context from each of individual reviewer abstracts.  
Also, we have not found strong evidence that such a low-rank approximation necessarily results in a more meaningful topic than the constituent word-embeddings themselves, which diminishes the explainability of the recommendations.  

Inspired by conference chairs and journal editors, graph-based models were proposed \cite{RWR,graphbased1} that use meta data such as co-authorship and citations to measure the relevance of the reviewer to a submission. 
We assume that meta-data is complementary to reviewer abstracts and can be used to further improve the outcome of any approach based on abstracts. 

Our proposed approach generates a co-occurrence graph for each of the reviewer's abstract. Each co-occurance graph has concepts as nodes which are present in reviewer's abstract and are also relevant to the submission. We will call this cooccurance graph as a clique and is used to build submission aware reviewer profile. For the purpose of simplicity, we define concept as a quality phrase which identifies an idea, method, or mechanism in the domain of interest. A reviewer abstract is relevant to the submission if the concepts studied by the reviewer were conducted in the same local context as that of the submission. 
Moreover, the contribution of each concept to a reviewer's relevance is weighted and filtered based on the importance and role of concepts mentioned in the context of the submission abstract as well as all the reviewer's abstracts. 

While evaluating the proposed approach, we found previously generated datasets inadequate, as explained in Section \ref{sec:data}. To overcome the limitation, we have worked with the community to create a new dataset, as explained in Section \ref{sec:our_dataset}. In addition, we have incorporated the approach into an end-to-end conference reviewer recommender system that has been adopted by many top-tier conferences. To preserve the anonymity of this paper we will not disclose the name of our system until this paper is published. \\
Our main contributions are as follows:
\begin{itemize}
    \setlength\itemsep{-0.3em}
    \item We propose a novel approach based on clique-document matrix to build a submission-aware reviewer profile and produce more precise recommendations than other popular systems.
    \item Our approach is based on reasoning and evidence that is explainable to program committee chairs who often require justification for reviewer recommendations.
    \item Our approach has been integrated into a working end-to-end reviewer recommender system that has been adopted by multiple top tier conferences.
    \item Collaborating with a research community we introduce a new evaluation dataset for reviewer recommendar systems that addresses the severe shortcomings of previously created datasets.
\end{itemize}
\begin{figure*}
    \centering
    \includegraphics[width=0.8\textwidth]{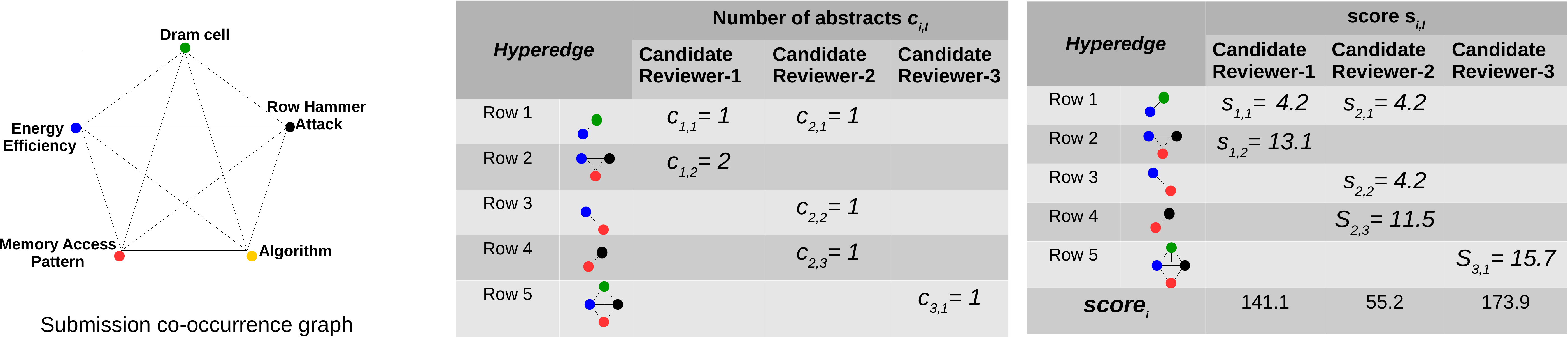}
    \hspace{4cm}(a)\hspace{4cm}(b)\hspace{4.9cm}(c)
    \caption{A running example illustrating the process of calculating relevance scores between an example submission and reviewers. Candidate reviewer-3 is recommended as the most relevant.}
    \vspace{-1.5em}
    \label{fig:example}
\end{figure*}

\section{Preliminaries}
This section identifies the key intuition and ideas that motivates the design of our approach.
\begin{itemize}
    \setlength\itemsep{-0.3em}
    \item A1: A reviewer's expertise can be modeled based on the abstracts they have published in the past, and thus treated as their profile.
    \item A2: The core ideas or concepts can be extracted from an abstract as (technically meaningful) phrases mentioned in the abstract.
    \item A3: To avoid dependence on specific usage of vocabulary, the set of concepts extracted from the submission should be extended with concepts similar in meaning for the domain. 
    \item A4: It is important to learn the local context in which a reviewer studied a concept and use it to qualify the relevance between the reviewer and a submission.
    \item A5: The role of each concept mentioned in the abstract should be determined and used to quantify the importance of the concept.
    \item A6: A reviewer with more conceptual overlap in individual abstracts is more relevant than one whose overlap spreads across multiple abstracts.
    \item A7: Evidence to support recommendations is necessary to make the recommendations explainable/accountable to program committees.
\end{itemize}

\section{Model}\label{sec:model}
To explain our model, we will use a running example based on one submission to illustrate the flow and the key steps of our method. We choose to use the computer architecture domain as an example to discuss our method because our curated dataset is from that domain. Noted that our method applies to other domains such as AI, NLP and biology etc.\\
\textbf{Extracting concepts and the local context:} Our model represents each reviewer abstract as a collection of the important phrases. The group of phrases extracted from the submission not only represent concepts studied in the submission but also the local context for each concept in the group. Phrases also vary in terms of how meaningful they are in calculating the relevance of submission-reviewer pair. Phrases such as ``Machine learning'' or ``Data Mining'' in the AI domain and ``Compiler'', ``Processor'' or ``Memory'' in the computer architecture domain, can help identify the high-level topics but are not specific enough for differentiating among candidate reviewers in the community. Other examples of frequently occurring, non-differentiating phrases include ``accuracy'', ``precision'' and ``loss function'' in the AI domain, ``power consumption'', ``energy efficiency'', and ``speed up'' in the computer architecture domain or ``significant improvement'' and ``novel approach'' in general. We consider these as less effective phrases in finding the reviewer's relevance to the submission. In comparison, some phrases represent the specific attributes, such as the problems or the methods studied in the paper. These specific phrases are more important for a precise characterization of the abstract. Therefore, phrases are weighted according to their importance (specificity), in line with assumption A5.

The extraction of phrases happens through an automated phrase extraction pipeline. It is part of our end-end system but not in this paper's scope. 
In Fig. \ref{fig:example}(a), \{``dram cell'', ``row hammer attack'', ``memory access pattern'', ``algorithm'', ``energy efficiency''\} are phrases extracted from the example submission abstract, forming a clique. In Fig\ref{fig:example}(b)-(c) each concept is represented as a colored circle.\\
\textbf{Replacement of Acronym Pairs:} We replace short-forms of any concept with its long-form using our acronym detection module (ex. graphics processing unit replaces GPU).
This reduces both repetition of equivalent phrases within an abstract as well as vocabulary mismatch between abstracts. \\
\begin{figure}[h!]
    \centering
    \includegraphics[width=.8\columnwidth]{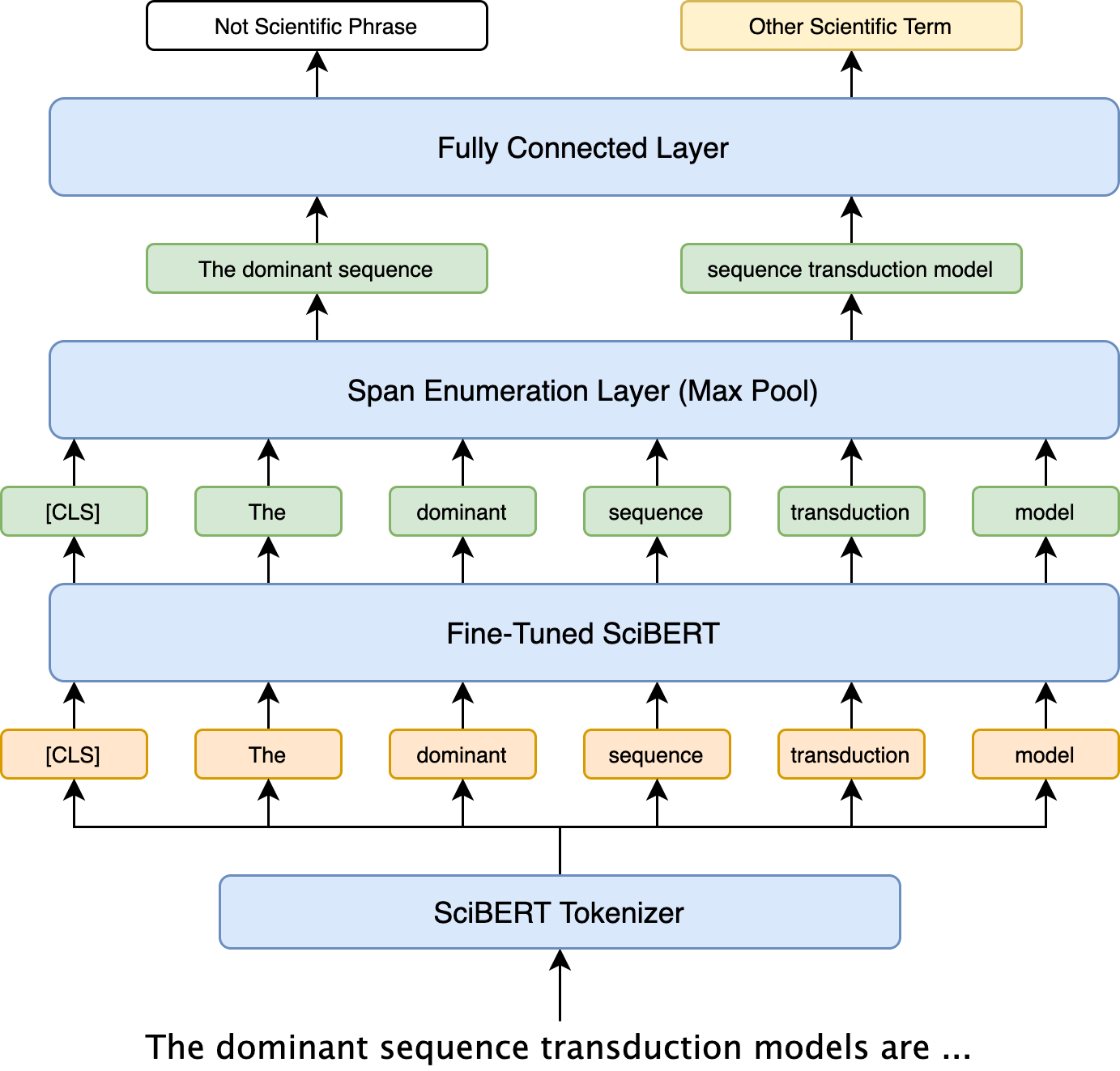}
    \caption{NER Model for Role Identification.  Tokens are shown in orange and embeddings are shown in green.  In the diagram, two example spans are shown while many more exist. The output space is a 7-dimensional vector corresponding to the 6 roles and a "Null" output for spans that are not a scientific phrase.}
    \vspace{-1em}
    \label{fig:ner}
\end{figure}\\
\textbf{Role Identification of Scientific Phrases:} Using the SciERC dataset \cite{scierc}, we trained a named entity recognition (NER) model aimed at classifying scientific phrases into six roles: Method, Task, Metric, Material, Other Scientific Term, and Generic.  We use a pretrained language model \cite{scibert} to obtain token embeddings, pool contiguous word spans using max pooling, and classify roles using a linear classifier as show in Fig. \ref{fig:ner}.  While previous works \cite{scierc, spert, dygiepp} have shown slight improvements while training on multi-task objectives, we opted to focus solely on NER due to runtime considerations and its relevance to our downstream tasks. The entity labels are used to filter scientific terms and determine which ones are most relevant to the meaning of the abstract. Phrases from irrelevant classes are not considered in further calculations. Figure \ref{fig:roles} shows how the classification and filtering of phrases improves the performance of the system. \\
\textbf{Reviewer's relevance to the submission:} We first calculate the relevance between the submission and each of the individual abstracts published by the reviewer.
From each of the reviewer's abstracts, we find edges shared with the submission co-occurrence graph (clique). 
The number of common edged between cliques obtained from reviewer and submission abstracts provides us a measure to quantify the degree to which a reviewer studied concepts in the context of each other in line with Assumption A4 and which are also relevant to the submission. 
Some of the partially or fully overlapped cliques with submission clique may appear in multiple reviewer abstracts. Let $G_{i}=\{g_{i,1}, g_{i,2}, ..., g_{i,l}\}$ be the set of all the $l$ unique intersecting cliques we found from the abstracts of the `$i^{th}$' reviewer. For example, Fig~\ref{fig:example}(b) shows that $G_1$, the set of cliques for Candidate Reviewer-1, has two unique cliques, partially overlapped with submission clique. The first clique $g_{1,1}$ consists of `dram cell' and `energy efficiency' and the second clique $g_{1,2}$ consists of `memory access pattern', `row hammer attack', and `energy efficiency'. 
Each overlapping clique from reviewer abstract contributes to the reviewer's profile in the submission context. These are then organized into a clique-document matrix as shown in figure \ref{fig:example}(b). 
With the clique-document matrix we explicitly capture concept specific relationships in an abstract which are missed altogether in the traditional approaches.

Each row in Fig. \ref{fig:example}(b) corresponds to one of the unique cliques from all the possible submission/reviewer-abstract pairs. Row 1 in Fig. \ref{fig:example}(b) shows that the candidate reviewer-2 studied two of the submission concepts `dram cell' and `energy efficiency' in each others context, while Row 3 shows that the same reviewer studied 'memory access pattern' and `energy efficiency' as a single context. Candidate reviewer-1 appears to have studied all the concepts as candidate reviewer-2, but reviewer-2 studied the two subgroups of concepts in isolation where as review-1 studied three of the concepts within the same paper, which makes reviewer-2 less relevant to the submission as compared to reviewer-1. Candidate reviewer-3 has only one abstract but covers four of the submission concepts in a single local context, creating a clique of size 4. Though this reviewer appears to be less experienced, he/she has written the most relevant abstract to the submission. The idea is that the reviewer who has studied the most number of the submission's concepts together in a paper has done the most similar study as the submission.

We define $C_{i,j}$ to track the number of reviewer-$i$'s abstracts in which the clique $g_{i,j}$ results from the intersection between a reviewer abstract and the submission.
For example, in Fig. \ref{fig:example}(b), the clique in the second row is the group of intersecting concepts between the submission and two of reviewer-1's abstracts. Intuitively, if a reviewer has studied the same group of concepts multiple times, they should be more qualified than someone who has studied the group of concepts fewer times. This number will be used to weigh the contribution of each clique to the relevance measure of a reviewer.

Let $N_{i,l}=\{n_{i,l}^{1}, n_{i,l}^{2}, ..., n_{i,l}^{k}\}$ be the concepts in $g_{i,l}$, and $W_{i,l}=\{w_{i,l}^{1}, w_{i,l}^{2}, ..., w_{i,l}^{k}\}$ be the set of weights corresponding to the concepts in $N_{i,l}$. The weights are used to assign importance to individual concepts in line with specificity as discussed earlier. To accomplish this goal, each $w_{i,l}^{j}$ is calculated using the \textit{tfidf} score corresponding to that concept \cite{tfidforiginal}. 
That is, the reviewer abstracts with more specific concepts relevant to the submission receive a higher score as compared to other abstracts with more generic concepts. For example, in Figure~\ref{fig:example}(c), reviewer-2's abstract with concepts ``row hammer attack'' and ``memory access pattern'' (Row 4) receives higher score than the abstracts with concept groups $\{$``energy efficiency'',``memory access pattern''$\}$ (Row 3) or $\{$``energy efficiency'', ``dram cell''$\}$ (Row 1).

To know the local context in which a concept was studied, there must be a minimum of two elements to find a clique. Thus, a reviewer abstract is not considered if $|N_{i,l}|<2$, where $|N_{i,l}|$ is the number of nodes in the clique. For example, the size of the clique in Row 2  in Fig. \ref{fig:example}(b) is three, whereas the size of the one in row five is four. In the rest of the rows in Fig. \ref{fig:example}(b), the size of each clique is two. The value of $|N_{i,l}|$ is used in the calculation of the relevance score of candidate reviewer-$i$ as explained below.
Intuitively, $|N_{i,l}|$ is the measure of co-occurrence of concepts in our model.

In order to calculate the reviewer's relevance to the submission, first a score is calculated corresponding to each of the cliques in $G_{i}$, as shown in Fig. \ref{fig:example}(c). The score for $g_{i,l}$ depends on the number of elements in $N_{i,l}$ which is also the number of nodes in a clique and their importance. Let $s_{i,l}$ be the score corresponding to $g_{i,l} \in G_{i}$. It is calculated as
\begin{equation}
    s_{i,l} = |N_{i,l}|^2 * \sum_{n_{i,l}^j \in N_{i,l}}w_{i,l}^j
\end{equation}
The first term $|N_{i,l}|^2$ scales the score according to the number of concepts in $g_{i,l}$. In other words, a reviewer abstract with larger overlap with the submission is scored much higher. This emphasizes the cohesion of the coverage of concepts by each reviewer. Without this scaling factor, for example, Reviewer-2 in Fig. \ref{fig:example}(b) would become more relevant to the submission compared to Reviewer-3. The final relevance between the reviewer and the submission is calculated as 
\begin{equation}
    score_{i} = \sum_{1}^{l}\{ln(1 + c_{i,l}) * s_{i,l}\}
\end{equation}
Equation 2 takes the sum of scores corresponding to all the unique cliques. The logarithm of the term $c_{i,l}$ de-emphasizes repeated use of the same clique. Otherwise, reviewers with more frequent publications may become more relevant even without any of their abstracts being more relevant to the submission. Thus, in line with preliminary A6, stronger matches are those that match more concepts in a single local context which is an intuitive measure of relevance that reflects human reasoning about reviewer similarity. The final scores for our example are shown in the last row of table in Fig. \ref{fig:example}(c). Reviewers are then ranked based on this final score.
\vspace{-2em}
\section{Dataset}
\subsection{Existing Evaluation Datasets}\label{sec:data} 
Prior evaluation datasets \cite{Mimno2007, RWR} have some severe limitations. For the dataset used in \cite{Mimno2007}, labels were incomplete and ascertained only for the top recommendations of the proposed system in \cite{Mimno2007}. The dataset consists of 33 NIPS papers accepted in 2006 and abstracts from the publications of 364  potential reviewers. There are only 330 labels in total, less than 10\% out of the $33 \times364$ total possible labels. On a scale from 0-3 the labels were generated by 9 annotators without confirmation by the potential reviewers, \cite{Mimno2007} which is another limitation and which is to say that labels are not "gold labels", making a fair comparison with new systems difficult. 
For another dataset, again pseudo-labels are created, which are not generated by the reviewers themselves \cite{RWR}. Instead, the dataset creates a 25 dimensional topic vector for each reviewer and submission that is used to infer the pseudo-labels. Each of 25 dimension is a high-level topic. Absence of gold labels from the reviewers themselves and system specific labels limits the trustworthiness of the labels and hinders proper evaluation.

\subsection{Proposed Evaluation Dataset}\label{sec:our_dataset}
In collaboration with research community, we have assembled a new dataset for evaluation purposes and available at \cite{goldlabels}. The dataset treats 10 previously published abstracts from computer architecture domain as submissions. The abstracts were selected by domain experts to cover a representative range of topics. There are 103 reviewers in the dataset who have served on the program committees of top-tier conferences in the domain. With > 1000 labels it is the largest test data for this application to our knowledge. We contacted the reviewers by email and requested them to score their \textit{``relevance''} and \textit{``interest''} on a scale from 0-3 for each of the submission abstracts.
Score of 0 corresponds to \textit{`not relevant'} or \textit{`not interested'}, while 3 corresponds to \textit{`highly relevant'} or \textit{`highly interested'}. In this work, we only use labels for relevance. However, for future work, we think that labels for interest will be useful. 

\vspace{-1em}
\section{Experiments and Results}
\subsection{Evaluation Metric}
 For performance evaluation, we use precision@k (P@k) which is defined to be the percentage of relevant reviewers in the top-k recommendations. If a reviewer with missing labels happen to be in the top-k, we ignore that recommendation from our precision calculations for any of our baselines and the proposed approach. For our curated dataset, we also evaluate using the average score@k (AS@K) which complements P@K by taking into account the average strength of the recommendations.
 \subsection{Baselines}
 We directly compare with other open-source recommendar systems including the ones used by top tier NLP conferences. Furthermore, since the proposed approach relies only on the submissions and reviewer abstracts, it can be used as a drop in substitute for any document similarity measure currently employed. As a result, the baselines also include document similarity measures (Doc2Vec, WMD, LDA, Common Topics, BERT) which are widely used in other reviewer matching systems for matching reviewer abstracts to that of the submissions. For example, the popular Toronto Paper Matching System uses LDA to compute document similarity. So while we cannot replicate the full TPMS system, LDA is the isolated portion of their method which is comparable to ours. Moreover, we also include widely adopted deep learning models like BERT \cite{bert} that are often used for document representation.   
 \begin{itemize}
    \setlength\itemsep{-0.3em}
     \item Doc2Vec \cite{doc2vec} - is neural network pretrained model to generate submission and reviewer representations. Reviewer-submission relevance is quantified by the cosine similarity of their embeddings.
     \item WMD \cite{WMD} - is distance metric between two documents based on pretrained word-embeddings. 
     \item LDA \cite{LDA} - is one of the most popular approaches used in reviewer recommendation systems for submission and reviewer representation. This model represents a document as a mixture of topics where each topic is distribution over words.
     \item Common Topics \cite{pare} - uses canonical correlation to jointly find topic vectors from reviewer and submission abstracts.  
     \item Hidden Topics \cite{hiddentopic} - represents documents as hidden topic vectors learned by applying matrix decomposition on a matrix where each row or column is an embedding vector corresponding to a word present in the document.
     \item BERT \cite{bert}- is one of the most popular deep learning models to represent a document. We use a pretrained BERT model to encode each of the reviewer or submission abstract.
     \item ACL-Org \cite{wieting19simple} - Reviewer matching method used by ACL-2020, ACL-2021, and NAACL 2021 for reviewer matching. We use configurations as reported in their code \cite{acl-github}.
 \end{itemize}

\subsection{Dataset}\label{sec:RA}
\begin{figure}[h!]
    \centering
    \includegraphics[width=0.8\columnwidth]{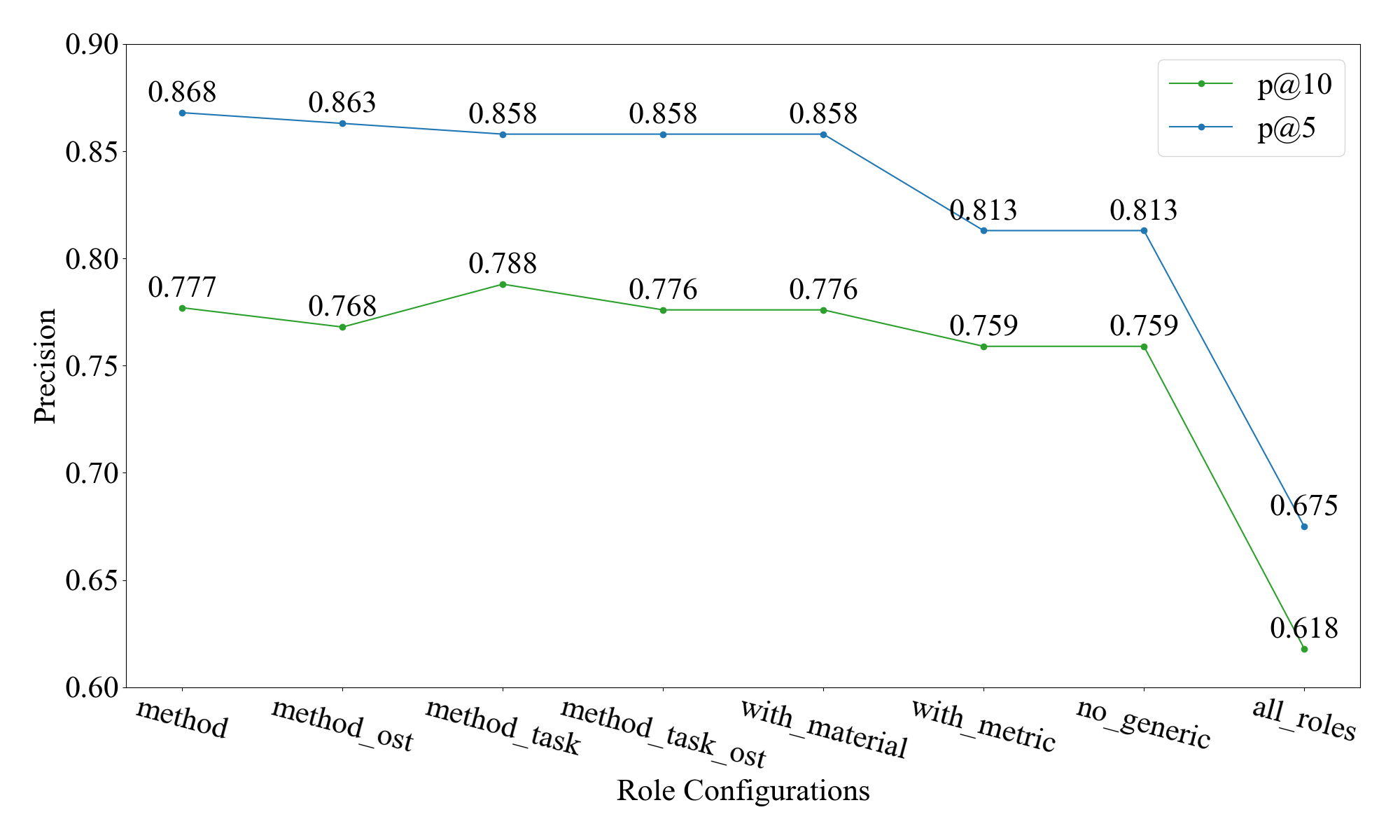}
    \vspace{-1em}
    \caption{Performance of Different Phrase Role Filtering Configurations. "with\_material" is the configuration including Methods, Tasks, Other Scientific Term (OST), and Materials.  Similarly, "with\_metric" is the configuration including Methods, Tasks, OST, and Metric.  "no\_generic" includes all roles except for Generic.}
    \vspace{-1em}
    \label{fig:roles}
\end{figure}

\begin{table}[t]
\centering
\begin{center}
\tiny
\begin{tabular}{|c||c|c|c|c|c|}
\hline
 \textbf{Method} &  \textbf{P@5} & \textbf{AS@5} & \textbf{P@10} & \textbf{AS@10}\\
 \hline
 \textbf{Doc2Vec} &  0.61 & 1.66 & 0.55 & 1.56\\
 \hline
 \textbf{WMD} &  0.13 & 0.61 & 0.14 & 0.69\\
 \hline
 \textbf{LDA} &  0.64 & 1.74 & 0.63 & 1.71\\ 
 \hline
 \textbf{Common Topics} &  0.52 & 1.48 & 0.50 & 1.50\\ 
 \hline
 \textbf{Hidden Topics (HT)} &  0.70 & 2.04 & 0.68 & 1.81\\ 
 \hline
 \textbf{BERT} &  0.54 & 1.53 & 0.47 & 1.47\\ 
 \hline
 \textbf{ACL-Org} &  0.63 & 1.78 & 0.64 & 1.73\\ 
 \hline
 \textbf{Proposed Approach} & \textbf{0.86} & \textbf{2.32} & \textbf{0.77} & \textbf{2.18}\\ 
 \hline
\end{tabular}
\end{center}
\caption{Results on our evaluation dataset. P@K and AS@K corresponds to the precision at K and average of annotated expertise level at K.}
\label{tab:results}
\end{table}
We demonstrate the proposed system on our newly collected dataset with gold labels. The results are shown in Table \ref{tab:results}. The proposed model improves the precision at 5 and precision at 10 by 16\% and 9\%, respectively compared to second best model among the baselines. In addition the proposed model achieves the best average score both at 5 and 10, illustrating the strength of the recommendations. We have also added an ablation study to show how phrase in different roles affect the performance, as shown in Fig. \ref{fig:roles}. 
Our systems have already been adopted by a number of top-tier conferences for reviewer recommendation, and the feedback on the quality of results were high. Here is an anonymized quote from one of the TPC chairs from one of the conferences we worked with: the proposed reviewer recommendation system ``saved us days of effort.  It also found more and better experts than we likely would have found otherwise.''
\begin{table}
    \centering
    \begin{center}
    \tiny
    \begin{tabular}{|p{.14\linewidth}|p{.80\linewidth}|}
        \hline
        \textbf{Summary of original abstract for a randomly selected test submission \cite{testsubmission}} & In our test submission, data and intratile parallelism in the winograd transformation algorithm is exploited to accelerate the training of convolution layer in convolutional neural networks. The algorithm inherently has more communication overhead due to large number of data accesses. The authors propose 3D stacked memory for near-data processing to minimize the data access cost. Furthermore, a reconfigurable network channel connectivity is proposed to balance the communication between workers processing a convolution layer. The work is compared against near data processing and a multi-GPU  system as a baseline.\\
         \hline
    \end{tabular}
    \end{center}
    \vspace{-1em}
    \caption{Summary of Example Abstract}
    \vspace{-1.5em}
    \label{tab:abstract}
\end{table}

\subsection{Qualitative Analysis}
In addition, we perform a qualitative analysis of the results to understand the behavior of our system to guide future directions for research. To add clarity, we sample a random submission from the collected dataset and analyze some of the top recommendations by different approaches. The sample submission abstract is presented in table \ref{tab:abstract}, and the recommended reviewers for the top 3 performing systems are given in table \ref{tab:sample_results}.

\textbf{Proposed Approach:} 
Out of the top 3 reviewers, 2 of the reviewers our system recommends are relevant. The top recommended reviewer, Reviewer 19, has written several papers in the same area as the submission. They also are an author of the submission. While that reviewer would be ineligible to review the paper, it is very likely they are the most qualified reviewer in the dataset, so it provides a good validity check that our model identified them. For the incorrect reviewer, reviewer 47, the system suggests that their most relevant paper to the submission was, ``Transparent Offloading and Map-ping (TOM) Enabling Programmer-Transparent Near Data Processing in GPU systems''. The phrases shared between that paper and the submission were: ``near data processing'', ``stacked memory'', ``data'', ``graphics processing unit'', and ``memory''. In this case, the intersecting phrases were misleading of the topic of the paper, focusing too much on hardware and memory but not capturing that the sample submission only considered memory in the context of neural network training.

\textbf{Hidden Topic Model:} Similar to the proposed approach, hidden topic model correctly recommended Reviewer 19, and incorrectly recommended reviewer 47. Note that the order of the two is reversed, compared to the proposed approach, suggesting that by creating the reviewer profile without knowledge of the context of the submission, hidden topics obscured the author's topics that were most similar to the submission. Contrary to our proposed approach, it is difficult to say what exactly contributed to reviewer 47 being chosen. 
To gain a better understanding of the reviewer match, we can look at the words with embeddings closest to the identified hidden topics. For the first and second hidden topics closest words were found to be ``young'', ``famous'', and ``man'' and ``gpu'',``gpus'' and ``cuda'', respectively. There is an apparent lack of reasons why the first topic is relevant. Moreover, second topic with a very specific focus on GPUs is not the prevalent topic. Topics are not representing any information about the local context which other topics may provide but fuzziness in the other learned topics is limiting the performance. 

\textbf{LDA:} Of the 3 reviewers suggested by LDA, 1 is labeled as relevant. Absent from the recommendations is Reviewer 19, one of the authors of the paper. The reason is the topics learned by LDA are high level 
topics from the entire domain, missing the more specific context in which a particular reviewer studied those topics. So the topics learned are too diluted to identify the relevance of Reviewer 19 to the submission. Moving to Reviewer 27, the first false positive recommended by LDA, the true research focus for this reviewer is programming languages and programming models \cite{adrian}. The reviewer does have papers intersecting approximate computing and computer vision which leverages neural networks so perhaps this similarity is responsible for the latent topic that creates the match. This is one disadvantage of LDA: topics are analyzed only in general contexts and again interpretation of results is difficult. For Reviewer 45, it appears that the topics identified obscured the specific research of the reviewer in a similar fashion. 
\begin{table}[t]
    \centering
    \begin{center}
    \tiny
    \begin{tabular}{|c|c|c|c|c|c|c|}
    \hline
     & \multicolumn{2}{l|}{\begin{tabular}[c]{@{}l@{}}Proposed\\ Approach\end{tabular}} & \multicolumn{2}{|l|}{\begin{tabular}[c]{@{}l@{}}Hidden\\ Topic\end{tabular}} & \multicolumn{2}{|l|}{LDA}\\
        \hline
        Rank & ID & Label & ID & Label & ID & Label  \\
        \hline
        1 & 19 & 3 & 92 & 3 & 40 & 2 \\
        2 & 47 & 1 & 47 & 1 & 27 & 1 \\
        3 & 69 & 3 & 19 & 3 & 45 & 1 \\
    \hline
    \end{tabular}
    \end{center}
    \vspace{-1em}
    \caption{Results from our randomly sampled abstract}
    \vspace{-1.5em}
    \label{tab:sample_results}
\end{table}

\subsection{Semantic Extension}
A current weakness in this method of matching is the dependence on specific language. To alleviate this problem, we attempted to use word embeddings to detect cliques with similar semantic meaning. For this work we relied on a relatively simple word2vec model to create domain specific embeddings \cite{w2v}. The cliques were then extended to consider very similar word vectors as matches.

While extending concepts with similar phrases, we noticed a reduction in precision due to information shift. For example, we found phrases such as "memory access" and "gpu" being extended with "cache access" and "cpu" respectively. However, these are different areas of expertise and top reviewers may end up to be from both the areas. Due to this, the precision at 5 and 10 with the semantic extension dropped to .82 and 0.71 respectively because of this information loss. For this reason, we chose to exclude the use of these embeddings in our final system but defer further exploration to future work.
\vspace{-1em}
\section{Related work}

\textbf{Keyword Based Models:} These methods rely on automatic keyword extraction using \textit{tfidf} based scores \cite{tfidf1,tfidf2,tfidf3}. These methods miss the specific context of the submission.

\textbf{Probabilistic Topic Models:} Most commonly used model is Latent Dirichlet Allocation (LDA) \cite{LDA} which represents topics as distribution over the words and document as a distribution over the topics. LDA assumes that the words are generated independently \cite{LDA2} which limits the ability of the topic model to capture the semantic relations between the words. Moreover, the words in a phrase may get associated to different topics leading to an incoherent topic representation \cite{phtm1,phtm2,phtm3}. LDA requires a very large corpus and generally captures the topics in larger context of the corpus. More specific reviewer expertise is not visible within these topics. The well known Toronto Paper Matching System (TPMS) \cite{tpms} is also based on LDA for topic modeling.

\textbf{Vector Embedding Models:} LSI \cite{LSI} is a commonly used model to represent documents as vectors which uses singular value decomposition (SVD) \cite{svd} for low rank approximation. INFOCOM reviewer assignment system \cite{infocom} presents a comparison between LSI and LDA. LSI is also able to capture the semantic relations between the words. However, similar to LDA, it requires large corpus to effectively capture these relations \cite{infocom}. Thus, vectors from LSI represent a reviewer in a global perspective rather than highlighting the reviewer's specific expertise relevant to a particular submission. 

Another extension to LSI based models is to use word embeddings rather than the words as input to SVD as in \cite{pare}. SVD can be computed independently for each reviewer where the matrix is constructed using word embeddings for the words appearing in the reviewer's abstracts. It assumes that the low-rank approximation are representative of the underlying topics in a reviewer profile. It is unclear what relations between hypothetical dimensions of different word embeddings are captured in the low-rank approximation.

\textbf{Word's Mover's Distance (WMD)} \cite{WMD} use word embeddings as input and formulates the distance between two documents as an optimal transport problem. It measures the cost of moving words from one document to another in the embedding space.

\textbf{Neural Network Models}
Neural network models have been used to learn the vector representation of documents \cite{nn1,nn2,nn3,nn4,doc2vec}. The relevance of reviewer to the submission is usually calculated as some measure of similarity between the vectors, the most commonly used measure being cosine similarity.

\textbf{Graph Models}
These models rely on meta information such as coauthors, venue and citations extracted from the publications. \cite{RWR} use co-author network, with weights of the edges indicating the number of papers in co-authorship.
We consider this meta information as complimentary information which can be used in addition to the content of the document. Since our focus in this paper is relevance matching based on the content, we will leave it for the future work.     
\section{Future Work}
For future work, we point to new, synergistic research directions that can further improve the results.
\begin{itemize}
    \setlength\itemsep{-0.3em}
    \item Explore more effective methods to capture concepts which are semantically equivalent without changing the domain of expertise.
    \item Use meta-data for more informed recommendations
\end{itemize} 
\section{Conclusion}
In summary, we propose an automated approach to reviewer recommendation that relies on a reviewer's previously published work. 
This work extracts explicit, interpretable concepts instead of ambiguous topic vectors or distributions. By considering these concepts and their co-occuring concept group as a clique and formulating the clique-document matrix, the system can capture the local context in which a reviewer studied a concept. This proposed approach shows an improved precision and explainability of results over the existing methods, illustrating the importance of local contexts. Furthermore, we also explained the procedure we followed to build a new dataset to over come the limitations of prior datasets.

\bibliography{anthology,custom}

\begin{thebibliography}{39}
\expandafter\ifx\csname natexlab\endcsname\relax\def\natexlab#1{#1}\fi

\bibitem[{acl()}]{acl-github}

\newblock Acl reviewer matching code.
\newblock \url{https://github.com/acl-org/reviewer-paper-matching}.

\bibitem[{gol()}]{goldlabels}

\newblock Gold-labels.
\newblock \url{https://github.com/oa98105/revscope-data}.

\bibitem[{acl(accessed November 23, 2020)}]{acl_acceptance}
 accessed November 23, 2020.
\newblock \href
  {https://aclweb.org/aclwiki/Conference_acceptance_rates#Main_Session_-_long_papers}
  {\emph{Conference acceptance rates}}.

\bibitem[{ACL(2019)}]{aclBlog}
PC~Chairs ACL. 2019.
\newblock What’s new, different and challenging in acl 2019?
\newblock \url{http://acl2019pcblog.fileli.unipi.it/?p=156}.
\newblock Accessed: 2019-05-19.

\bibitem[{Anjum et~al.(2019)Anjum, Gong, Bhat, Hwu, and Xiong}]{pare}
Omer Anjum, Hongyu Gong, Suma Bhat, Wen-Mei Hwu, and Jinjun Xiong. 2019.
\newblock \href {https://doi.org/10.18653/v1/D19-1049} {Pare: A paper-reviewer
  matching approach using a common topic space}.
\newblock pages 518--528.

\bibitem[{Beltagy et~al.(2019)Beltagy, Lo, and Cohan}]{scibert}
Iz~Beltagy, Kyle Lo, and Arman Cohan. 2019.
\newblock Scibert: A pretrained language model for scientific text.

\bibitem[{Blei et~al.(2003)Blei, Ng, and Jordan}]{LDA}
David~M. Blei, Andrew~Y. Ng, and Michael~I. Jordan. 2003.
\newblock \href {http://dl.acm.org/citation.cfm?id=944919.944937} {Latent
  dirichlet allocation}.
\newblock \emph{The Journal of Machine Learning Research}, 3:993--1022.

\bibitem[{Caines(2020, accessed August 30, 2020))}]{subtrends}
Andrew Caines. 2020, accessed August 30, 2020).
\newblock \href
  {http://www.marekrei.com/blog/geographic-diversity-of-nlp-conferences/}
  {\emph{conference publication trends}}.

\bibitem[{Deerwester et~al.(1990)Deerwester, Dumais, Furnas, Landauer, and
  Harshman}]{LSI}
Scott Deerwester, Susan~T Dumais, George~W Furnas, Thomas~K Landauer, and
  Richard Harshman. 1990.
\newblock Indexing by latent semantic analysis.
\newblock \emph{Journal of the American society for information science},
  41(6):391--407.

\bibitem[{Devlin et~al.(2019)Devlin, Chang, Lee, and Toutanova}]{bert}
J.~Devlin, Ming-Wei Chang, Kenton Lee, and Kristina Toutanova. 2019.
\newblock Bert: Pre-training of deep bidirectional transformers for language
  understanding.
\newblock In \emph{NAACL-HLT}.

\bibitem[{Eberts and Ulges(2019)}]{spert}
Markus Eberts and Adrian Ulges. 2019.
\newblock Span-based joint entity and relation extraction with transformer
  pre-training.

\bibitem[{El-Kishky et~al.(2014)El-Kishky, Song, Wang, Voss, and Han}]{phtm1}
Ahmed El-Kishky, Yanglei Song, Chi Wang, Clare~R. Voss, and Jiawei Han. 2014.
\newblock \href {https://doi.org/10.14778/2735508.2735519} {Scalable topical
  phrase mining from text corpora}.
\newblock \emph{Proc. VLDB Endow.}, 8(3):305–316.

\bibitem[{Falsafi et~al.(2018)Falsafi, Drumond, and Sutherland}]{sigarchBlog}
Babak Falsafi, Mario Drumond, and Mark Sutherland. 2018.
\newblock Isca'18 review process reflections.
\newblock \url{https://www.sigarch.org/isca18-review-process-reflections/}.
\newblock Accessed: 2019-05-19.

\bibitem[{Gong et~al.(2018)Gong, Sakakini, Bhat, and Xiong}]{hiddentopic}
Hongyu Gong, Tarek Sakakini, Suma Bhat, and Jinjun Xiong. 2018.
\newblock Document similarity for texts of varying lengths via hidden topics.
\newblock In \emph{Proceedings of the 56th Annual Meeting of the Association
  for Computational Linguistics (Volume 1: Long Papers)}, volume~1, pages
  2341--2351.

\bibitem[{He(2016)}]{phtm3}
Yulan He. 2016.
\newblock Extracting topical phrases from clinical documents.
\newblock In \emph{Proceedings of the Thirtieth AAAI Conference on Artificial
  Intelligence}, AAAI’16, page 2957–2963. AAAI Press.

\bibitem[{{Hong} et~al.(2018){Hong}, {Ro}, and {Kim}}]{testsubmission}
B.~{Hong}, Y.~{Ro}, and J.~{Kim}. 2018.
\newblock Multi-dimensional parallel training of winograd layer on
  memory-centric architecture.
\newblock In \emph{2018 51st Annual IEEE/ACM International Symposium on
  Microarchitecture (MICRO)}, pages 682--695.

\bibitem[{Jin et~al.(2018)Jin, Geng, Mou, and Chen}]{tfidf1}
Jian Jin, Qian Geng, Haikun Mou, and Chong Chen. 2018.
\newblock Author-subject-topic model for reviewer recommendation.
\newblock \emph{Journal of Information Science}, 45(4):554--570.

\bibitem[{Jones(1972)}]{tfidforiginal}
Karen~Spärck Jones. 1972.
\newblock A statistical interpretation of term specificity and its application
  in retrieval.
\newblock \emph{Journal of Documentation}, 28:11--21.

\bibitem[{Kawamae(2014)}]{phtm2}
Noriaki Kawamae. 2014.
\newblock \href {https://doi.org/10.1145/2556195.2559895} {Supervised n-gram
  topic model}.
\newblock In \emph{Proceedings of the 7th ACM International Conference on Web
  Search and Data Mining}, WSDM ’14, page 473–482, New York, NY, USA.
  Association for Computing Machinery.

\bibitem[{{Klema} and {Laub}(1980)}]{svd}
V.~{Klema} and A.~{Laub}. 1980.
\newblock The singular value decomposition: Its computation and some
  applications.
\newblock \emph{IEEE Transactions on Automatic Control}, 25(2):164--176.

\bibitem[{Kusner et~al.(2015)Kusner, Sun, Kolkin, and Weinberger}]{WMD}
Matt~J. Kusner, Yu~Sun, Nicholas~I. Kolkin, and Kilian~Q. Weinberger. 2015.
\newblock From word embeddings to document distances.
\newblock In \emph{Proceedings of the 32nd International Conference on
  International Conference on Machine Learning - Volume 37}.

\bibitem[{Lau and Baldwin(2016)}]{nn4}
Jey~Han Lau and Timothy Baldwin. 2016.
\newblock An empirical evaluation of doc2vec with practical insights into
  document embedding generation.
\newblock \emph{arXiv preprint arXiv:1607.05368}.

\bibitem[{Laurent and Zemel(2013)}]{tpms}
Charlin Laurent and Richard~S. Zemel. 2013.
\newblock The toronto paper matching system: An automated paper-reviewer
  assignment system.
\newblock In \emph{Proceedings of 30th International Conference on Machine
  Learning}.

\bibitem[{Le and Mikolov(2014{\natexlab{a}})}]{doc2vec}
Quoc Le and Tomas Mikolov. 2014{\natexlab{a}}.
\newblock Distributed representations of sentences and documents.
\newblock In \emph{International conference on machine learning}, pages
  1188--1196.

\bibitem[{Le and Mikolov(2014{\natexlab{b}})}]{nn1}
Quoc Le and Tomas Mikolov. 2014{\natexlab{b}}.
\newblock Distributed representations of sentences and documents.
\newblock In \emph{International conference on machine learning}, pages
  1188--1196.

\bibitem[{Li and Hou(2016)}]{infocom}
Baochun Li and Y~Thomas Hou. 2016.
\newblock The new automated ieee infocom review assignment system.
\newblock \emph{IEEE Network}, 30(5):18--24.

\bibitem[{Lin et~al.(2015)Lin, Liu, Yang, Li, Zhou, and Li}]{nn3}
Rui Lin, Shujie Liu, Muyun Yang, Mu~Li, Ming Zhou, and Sheng Li. 2015.
\newblock Hierarchical recurrent neural network for document modeling.
\newblock In \emph{Proceedings of the 2015 Conference on Empirical Methods in
  Natural Language Processing}, pages 899--907.

\bibitem[{Liu et~al.(2014)Liu, Suel, and Memon}]{RWR}
Xiang Liu, Torsten Suel, and Nasir Memon. 2014.
\newblock A robust model for paper reviewer assignment.
\newblock In \emph{Proceedings of the 8th ACM Conference on Recommender
  Systems}, RecSys '14, pages 25--32. ACM.

\bibitem[{Luan et~al.(2018)Luan, He, Ostendorf, and Hajishirzi}]{scierc}
Yi~Luan, Luheng He, Mari Ostendorf, and Hannaneh Hajishirzi. 2018.
\newblock Multi-task identification of entities, relations, and coreferencefor
  scientific knowledge graph construction.

\bibitem[{Mikolov et~al.(2013)Mikolov, Chen, Corrado, and Dean}]{w2v}
Tomas Mikolov, Kai Chen, Greg Corrado, and Jeffrey Dean. 2013.
\newblock Efficient estimation of word representations in vector space.
\newblock \emph{arXiv: 1301.3781}.

\bibitem[{Mimno and McCallum(2007)}]{Mimno2007}
David Mimno and Andrew McCallum. 2007.
\newblock \href {https://doi.org/10.1145/1281192.1281247} {Expertise modeling
  for matching papers with reviewers}.
\newblock In \emph{Proceedings of the 13th ACM SIGKDD International Conference
  on Knowledge Discovery and Data Mining}, KDD '07, pages 500--509, New York,
  NY, USA. ACM.

\bibitem[{Nguyen et~al.(2018)Nguyen, Snchez-Hernndez, Agell, Rovira, and
  Angulo}]{tfidf3}
Jennifer Nguyen, Germn Snchez-Hernndez, Nria Agell, Xari Rovira, and Cecilio
  Angulo. 2018.
\newblock \href {https://doi.org/10.1016/j.patrec.2017.09.020} {A decision
  support tool using order weighted averaging for conference review
  assignment}.
\newblock \emph{Pattern Recogn. Lett.}, 105(C):114--120.

\bibitem[{Rodriguez and Bollen(2008)}]{graphbased1}
Marko~A. Rodriguez and Johan Bollen. 2008.
\newblock An algorithm to determine peer-reviewers.
\newblock CIKM '08, page 319–328. Association for Computing Machinery.

\bibitem[{Sampson(accessed August 30, 2020)}]{adrian}
Adrian Sampson. accessed August 30, 2020.
\newblock \href {https://www.cs.cornell.edu/~asampson/} {[link]}.

\bibitem[{Socher et~al.(2011)Socher, Lin, Manning, and Ng}]{nn2}
Richard Socher, Cliff~C Lin, Chris Manning, and Andrew~Y Ng. 2011.
\newblock Parsing natural scenes and natural language with recursive neural
  networks.
\newblock In \emph{Proceedings of the 28th international conference on machine
  learning (ICML-11)}, pages 129--136.

\bibitem[{Tang et~al.(2010)Tang, Tang, and Tan}]{tfidf2}
Wenbin Tang, Jie Tang, and Chenhao Tan. 2010.
\newblock \href {https://doi.org/10.1109/WI-IAT.2010.133} {Expertise matching
  via constraint-based optimization}.
\newblock In \emph{Proceedings of the 2010 IEEE/WIC/ACM International
  Conference on Web Intelligence and Intelligent Agent Technology - Volume 01},
  WI-IAT '10, pages 34--41, Washington, DC, USA. IEEE Computer Society.

\bibitem[{Wadden et~al.(2019)Wadden, Wennberg, Luan, and Hajishirzi}]{dygiepp}
David Wadden, Ulme Wennberg, Yi~Luan, and Hannaneh Hajishirzi. 2019.
\newblock Entity, relation, and event extraction with contextualized span
  representations.

\bibitem[{Wieting et~al.(2019)Wieting, Gimpel, Neubig, and
  Berg-Kirkpatrick}]{wieting19simple}
John Wieting, Kevin Gimpel, Graham Neubig, and Taylor Berg-Kirkpatrick. 2019.
\newblock \href {https://arxiv.org/abs/1909.13872} {Simple and effective
  paraphrastic similarity from parallel translations}.
\newblock In \emph{Proceedings of the Association for Computational
  Linguistics}.

\bibitem[{Xie et~al.(2015)Xie, Yang, and Xing}]{LDA2}
Pengtao Xie, Diyi Yang, and Eric Xing. 2015.
\newblock Incorporating word correlation knowledge into topic modeling.
\newblock In \emph{Proceedings of the 2015 conference of the north American
  chapter of the association for computational linguistics: human language
  technologies}, pages 725--734.

\end{thebibliography}
\bibliographystyle{acl_natbib}



\end{document}